\documentclass[%
 reprint,
superscriptaddress,
 amsmath,amssymb,
 aps,
prl,
]{revtex4-1}

\usepackage{graphicx}
\usepackage{dcolumn}
\usepackage{bm}
\usepackage{todonotes}
\usepackage{simplewick}
\usepackage{hyperref}
\hypersetup{
     colorlinks,
     linkcolor={blue!50!black},
     citecolor={blue!50!black},
     urlcolor={blue!80!black}
}
\graphicspath{{./figures/}}

\usepackage{xspace}
\makeatletter
\DeclareRobustCommand\onedot{\futurelet\@let@token\@onedot}
\def\@onedot{\ifx\@let@token.\else.\null\fi\xspace}

\newcommand{\etal}{{\it et al}\onedot}
\newcommand{\ie}{{i.e}\onedot}
\newcommand{\eg}{{e.g}\onedot}

\makeatother

\newcommand{\fmi}{\ensuremath{\,\text{fm}^{-1}}}

\newcommand{\MeV}{\ensuremath{\,\text{MeV}}}

\begin{document}

\title{A nucleus-dependent valence-space approach to nuclear structure}%
\author{S.~R. Stroberg}%
\email{sstroberg@triumf.ca}%
\affiliation{TRIUMF 4004 Wesbrook Mall, Vancouver BC V6T 2A3 Canada}%
\author{A. Calci}%
  \affiliation{TRIUMF 4004 Wesbrook Mall, Vancouver BC V6T 2A3 Canada}%
\author{H. Hergert}%
\affiliation{Facility for Rare Isotope Beams and Department of Physics and Astronomy, Michigan State University, East Lansing, MI 48844, USA }%
\author{J.~D. Holt}%
\affiliation{TRIUMF 4004 Wesbrook Mall, Vancouver BC V6T 2A3 Canada}%
\author{S.~K. Bogner}%
\affiliation{Facility for Rare Isotope Beams and Department of Physics and Astronomy, Michigan State University, East Lansing, MI 48844, USA }%
\author{R. Roth}%
\affiliation{Institute f\"ur Kerphysik, Technische Universit\"at Darmstadt 64289 Darmstadt, Germany}%
\author{A. Schwenk}%
\affiliation{Institute f\"ur Kerphysik, Technische Universit\"at Darmstadt 64289 Darmstadt, Germany}%
\affiliation{ExtreMe Matter Instistute EMMI, GSI Helmholtzzentrum f\"ur Schwerionenforschung GmbH, 64291 Darmstadt, Germany}%
\affiliation{Max-Planck-Institut f\"ur Kernphysik, Saupfercheckweg 1, 69117 Heidelberg, Germany}%

\begin{abstract}
We present a nucleus-dependent valence-space approach for calculating ground and excited states of nuclei, which generalizes the shell-model in-medium similarity renormalization group to an ensemble reference with fractionally filled orbitals. Because the ensemble is used only as a reference, and not to represent physical states, no symmetry restoration is required. This allows us to capture 3N forces among valence nucleons with a valence-space Hamiltonian specifically targeted to each nucleus of interest. Predicted ground-state energies from carbon through nickel agree with results of other large-space ab initio methods, generally to the 1\% level. In addition, we show that this new approach is required in order to obtain convergence for nuclei in the upper $p$ and $sd$ shells. Finally, we address the  $1^+_1$/$3^+_1$ inversion problem in $^{22}$Na and $^{46}$V. This approach extends the reach of ab initio nuclear structure calculations to essentially all light- and medium-mass nuclei.

\end{abstract}

\pacs{Valid PACS appear here}
\maketitle


The development of a first-principles, or ab initio, theoretical description of atomic nuclei is a central challenge in nuclear physics.
This task is complicated by the combined difficulties of not having an exact form for nuclear interactions and the great complexity in solving the nuclear many-body problem.
Regardless, controlled predictions with uncertainty estimates are vital to guide efforts of rare-isotope beam facilities \cite{Balantekin2014,Hebeler2015}, to constrain nucleosynthesis models predicting the origin of heavy elements in the universe \cite{Mumpower2016,Martin2016}, and to quantify nuclear structure effects in searches for beyond-standard-model physics such as neutrinoless double-beta decay \cite{Avignone2008}, dark matter \cite{Engel1992,Klos2013}, and superallowed beta decay \cite{Hardy2015}.
Developments in chiral effective field theory \cite{Epelbaum2009,Machleidt2011}, similarity renormalization group (SRG) \cite{Bogner2007}, and ab initio many-body techniques \cite{Carlson2015,Barrett2013,Hagen2014,Hergert2016,Dickhoff2004,Roth2009} provide a unified picture for these efforts,
while three-nucleon (3N) forces have emerged as an essential component of nuclear forces  \cite{Hebeler2015,Otsuka2010a,Hagen2012a,Cipollone2013,Hergert2013a,Holt2013a,Holt2012a,Hagen2012,Wienholtz2013,Hergert2014,Soma2014,Hagen2016,GarciaRuiz2016}.

One promising approach to the many-body problem is offered by the shell-model paradigm, where a valence-space Hamiltonian of tractable dimension is decoupled from the much larger Hilbert space and diagonalized.
This allows the treatment of excited states, deformation, and transitions in open-shell systems within a single framework.
Building upon earlier perturbative approaches \cite{Kuo1966,Hjorth-Jensen1995,Holt2005}, ab initio methods now provide shell-model Hamiltonians in a nonperturbative manner \cite{Lisetskiy2008,Tsukiyama2012,Bogner2014,Jansen2014,Dikmen2015,Stroberg2016,Jansen2016}, similar to recent work for chemical systems, see \eg, \cite{Yanai2006,*Yanai2007}.
However, the inclusion of residual 3N forces 
\footnote{Three-body forces will be induced by the decoupling of the valence space, even if they are not present in the initial Hamiltonian.}
among valence particles \cite{Caesar2013,Holt2014a} remains problematic in nonperturbative methods and leads to a loss in accuracy compared to large-space ab initio calculations \cite{Stroberg2016}.

A first attempt to address this shortcoming within the in-medium similarity renormalization group (IM-SRG) framework \cite{Stroberg2016} used normal ordering with respect to closed sub-shells in the valence space, but gave no clear prescription for systems far from closed shells.
In this Letter we generalize our approach to a reference with fractionally occupied orbits, capturing the dominant effects of neglected 3N forces among valence nucleons.
Since these effects scale non-trivially with mass number $A$, the standard shell-model approach of constructing one Hamiltonian for an entire region \cite{Brown2001a,Caurier2005} appears to be insufficient from an ab initio standpoint.
Therefore we adopt a new strategy, using the IM-SRG to decouple a targeted valence-space Hamiltonian for each nucleus, using a specialized reference for the normal ordering.
The resulting ground-state energies agree well with large-space ab initio methods, generally to the 1\% level.
We highlight the improvement for systems far from closed shells in $^{22}$Na and $^{46}$V, where the  $3^+_1/1^+_1$ level-inversion problem is addressed for the first time in an ab initio framework.

A key feature of the IM-SRG is the use of operators normal ordered with respect to a particular reference state $|\Phi_0\rangle$.
The Hamiltonian, which in free space has one-, two-, and three-body pieces, is rewritten exactly as \cite{Hergert2013a}
\begin{multline}
H = E_0 +  \sum_{ij}f_{ij}\{a_i^{\dagger}a_j\}
+ \frac{1}{4}\sum_{ijkl}\Gamma_{ijkl}\{a^{\dagger}_ia^{\dagger}_ja_la_k\}\\
+ \frac{1}{36}\sum_{ijklmn}W_{ijklmn} \{a^{\dagger}_ia^{\dagger}_ja^{\dagger}_ka_na_ma_l\},
\end{multline}
where the braces indicate normal ordering with respect to the reference,
such that $\langle\Phi_0|\{\mathcal{A}_1\ldots\mathcal{A}_N\}|\Phi_0\rangle\!=\!0$, where $\mathcal{A}$ represents either a creation or annihilation operator.
To make the calculation tractable, the residual 3N part $W_{ijklmn}$ is neglected, leading to the normal-ordered two-body approximation \cite{Hagen2007,Otsuka2010a,Roth2012}.
Naively, the quality of this approximation depends on how well $|\Phi_0\rangle$ approximates the eigenstate of the full Hamiltonian.

In the original implementation of the IM-SRG \cite{Tsukiyama2011,Hergert2013,Hergert2016}, $|\Phi_0\rangle$ was taken to be a single Slater determinant, limiting its applicability to ground states of closed-shell nuclei such as $^{16}$O.
The multi-reference formulation \cite{Hergert2013a,Hergert2014} extended the reach to general open-shell nuclei, though current implementations are limited to ground states of even-even nuclei.
%
%
The shell-model variant \cite{Tsukiyama2012,Bogner2014} of the IM-SRG uses the core of the valence space as the reference --- \eg, $^{16}$O for an $sd$ valence space -- enabling the treatment of open-shell nuclei and excited states.

Comparing the results of the above approaches provides a test of the validity of their respective approximations.
As was found in Refs.~\cite{Bogner2014,Stroberg2016}, the shell-model IM-SRG gives good agreement for few valence particles, but as more are added, the results become overbound relative to large-space methods.
This may be understood by considering that, for the choice of interaction used in those studies, the initial 3N forces combined with those induced by SRG and IM-SRG transformations produce a net repulsion
\footnote{For this Hamiltonian we find that the effect of the 3N force induced by the IM-SRG flow is typically of the same sign and comparable magnitude to that of the free-space SRG evolution.}.
The bulk of this repulsive 3N force is captured by the normal-ordered two-body approximation.
However, if the normal ordering is performed with respect to the core, 
the repulsive 3N interaction among valence nucleons is neglected, leading to overbinding that grows with the number of valence particles.

\begin{figure}
\includegraphics[width=1.0\columnwidth]{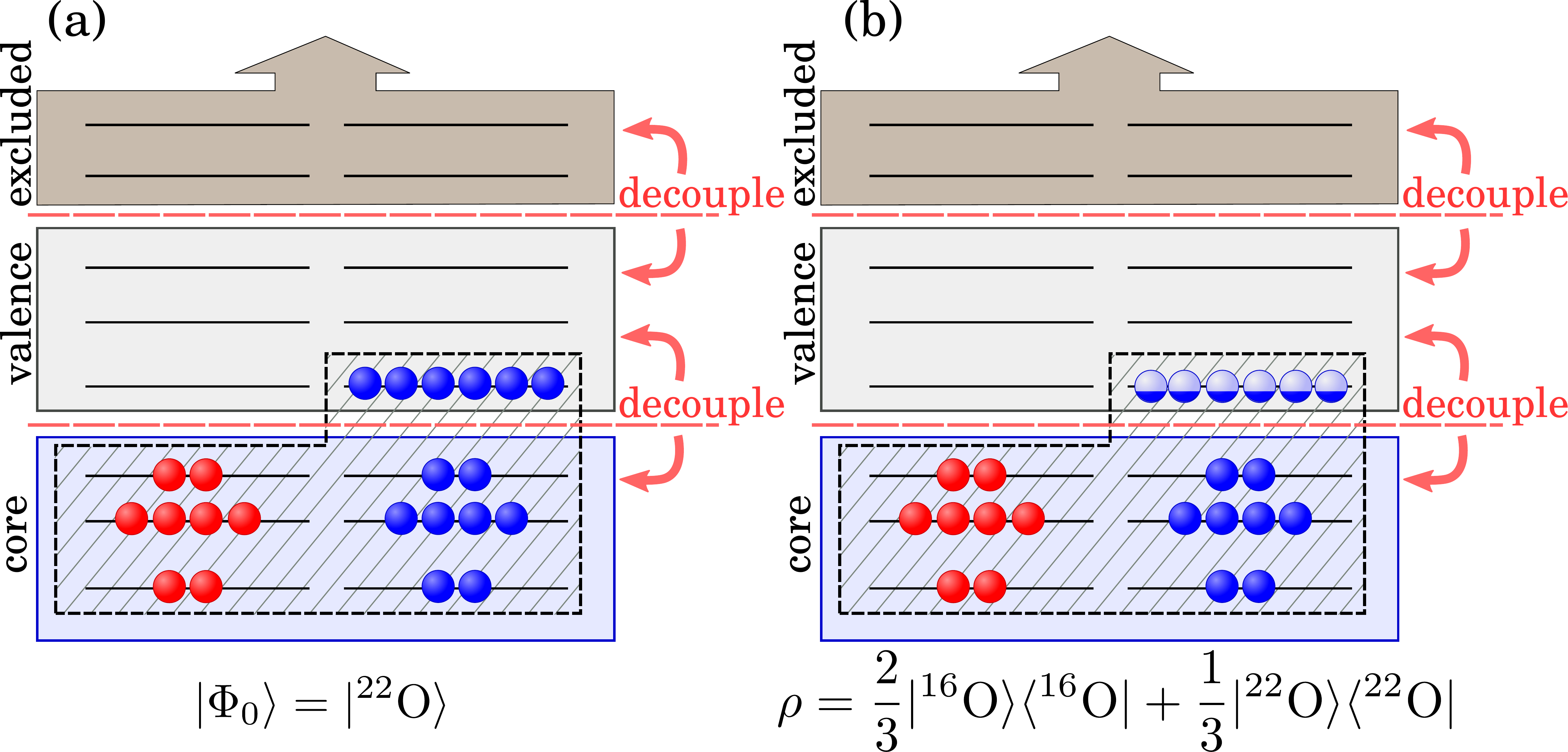}
\caption{Examples of (a) a reference which is different from the core, here $^{22}$O, and (b) an ensemble reference with fractional filling, here approximating $^{18}$O.}
\label{fig:ShellsIllustration}
\end{figure}

A first attempt to address this issue, illustrated schematically in Fig.~\ref{fig:ShellsIllustration}(a), is to normal order with respect to the nearest closed shell (\eg, using a $^{22}$O reference to calculate $^{23}$F), while still decoupling the core ~\cite{Stroberg2016}.
At the end of the decoupling, the interaction is re-normal-ordered with respect to the core for use in a standard shell-model code.
This final transformation, carried out at the two-body level, is unitary and preserves the decoupling.
This approach, referred to as targeted normal ordering (TNO), largely corrects the discrepancy between the shell-model results and other large-space methods for the oxygen and fluorine chains.

One caveat is that some doubly open-shell nuclei (\eg, $^{22}$Na)
are far from a closed shell.
A naive next step, illustrated in Fig.~\ref{fig:ShellsIllustration}(b), would be to ``interpolate'' between closed shells with a fractional occupation of the last shell ($0d_{5/2}$ in the case of $^{22}$Na),
ideally without the additional computational effort associated with the multi-reference formulation.
As discussed in Refs.~\cite{Duguet2001,Perez-Martin2008}, this so-called equal-filling approximation is equivalent to employing an ensemble or mixed-state reference composed of closed-shell Slater determinants of different particle number.
The reference is defined by a density matrix
$\rho = \sum_\alpha c_\alpha |\Phi_\alpha \rangle \langle \Phi_\alpha |$
with coefficients $c_\alpha$ chosen such that the ensemble-averaged particle number, expressed as a trace over the density matrix, is that of our system of interest, \ie, $\sum_p \mathrm{Tr}\left( \rho\, a^{\dagger}_pa_p \right)\!=\!A$.
Normal ordering is then defined such that $\mathrm{Tr}\left( \rho \, \{\mathcal{A}_1\ldots \mathcal{A}_N\} \right)\!=\!0$ \cite{Thouless1957,Gaudin1960}.
The Wick contraction is
\begin{equation}
\contraction{\{}{a}{}{a_q} \{ a^{\dagger}_p a_q \} = \sum_\alpha c_\alpha \langle \Phi_\alpha | a^{\dagger}_pa_q | \Phi_\alpha \rangle
\equiv n_p \delta_{pq},
\end{equation}
where $n_p$ is now the ensemble-averaged (fractional) occupation of the orbit $p$.
Similarly, $\contraction{\{}{a}{}{a^{\dagger}} \{ a_p a^{\dagger}_q\}=(1-n_p)\delta_{pq}$, and all other contractions vanish.
For closed shells, this ensemble normal ordering (ENO) reduces to the TNO described above.

\begin{figure*}
\includegraphics[width=2.0\columnwidth]{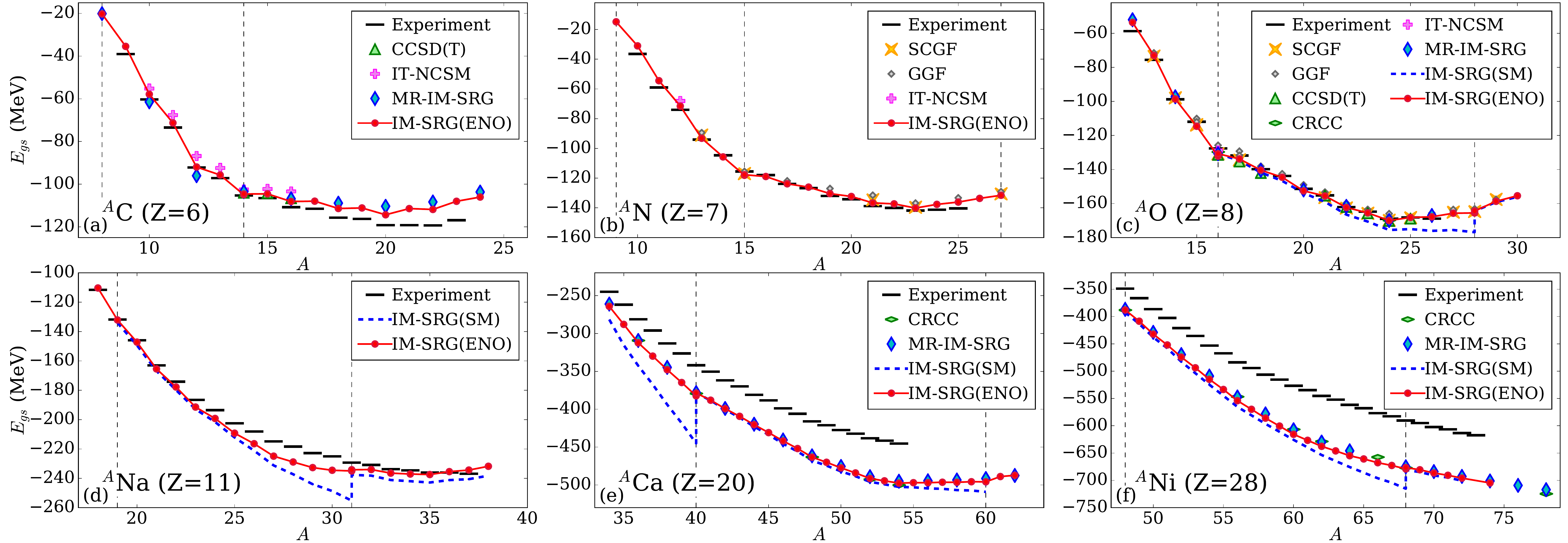}
\caption{Ground-state energies of medium-mass isotopic chains
calculated with multi-reference IM-SRG (MR-IM-SRG) \cite{Hergert2013a,Hergert2014}, coupled cluster (CCSD(T)) \cite{Jansen2014}, completely renormalized coupled cluster (CRCC) \cite{Binder2014}, importance-truncated no-core shell model (IT-NCSM, see text), self-consistent Green's function (SCGF) and Gor'kov Green's function(GGF) \cite{Cipollone2015}, compared to experiment \cite{Wang2012}. The IM-SRG(SM) curves use a core reference, while the curves labeled IM-SRG(ENO) use an ensemble reference. Dashed vertical lines indicate neutron harmonic-oscillator shell closures. We note calcium isotopes are calculated with $e_{\mathrm{max}}\!=\!14$, consistent with the MR-IMSRG calculations.}
\label{fig:GroundStates}
\end{figure*}

For ground-state methods (\eg, single-/multi-reference IM-SRG, coupled cluster, Hartree-Fock-Bogoliubov) such a reference is problematic because the resulting decoupled state is not a particle-number eigenstate and is therefore unphysical \cite{Perez-Martin2008}.
This is related to the problem of N-representability \cite{Mazziotti2007}.
However, in the valence-space approach the reference merely provides a convenient way to express operators; the Hamiltonian commutes with the particle-number operator throughout the calculation.
While the reference does not accurately represent any particular state in the targeted system, it is reasonable (and results confirm) that having the right number of particles in roughly the right configuration is a sufficiently good approximation.
Indeed, the exact eigenstate of the full Hamiltonian is not necessarily the optimal reference for making the normal-ordered two-body approximation \cite{Yanai2006}.
Gebrerufael \etal \cite{Gebrerufael2015} demonstrated that little is gained by improving the reference beyond a low-order approximation, and similar observations have been made in nuclear matter \cite{Carbone2014}.

Unless noted, the following calculations are carried out using the same interaction as in Ref.~\cite{Stroberg2016}, with $\lambda_{\mathrm{SRG}}\!=\!1.88\fmi$ \cite{Roth2014b} and an oscillator frequency of 24~MeV.
We employ a model space defined by all single-particle states with $2n+\ell\leq e_{\mathrm{max}}\!=\!12$.
The interaction is transformed to the Hartree-Fock basis, at which point the normal-ordered 3N interaction is discarded.
The valence space is defined by one major harmonic-oscillator shell for protons and neutrons (\eg, for $^{16}$N, protons in the $0p$ shell and neutrons in the $1s0d$ shell).
The IM-SRG decoupling is performed using the Magnus formulation presented in Ref.~\cite{Morris2015} with the arctangent variant of the White generator \cite{White2002}.
Valence-space diagonalizations are performed with NuShellX@MSU \cite{Brown2014}.
For comparison, in a few cases we also perform importance-truncated no-core shell model (IT-NCSM) \cite{Roth2009} calculations, fully including 3N forces and extrapolating to an infinite model space.


As a first test of this new approach, we investigate ground-state energies of medium-mass isotopic chains.
We emphasize that the target for success is not experimental data, but  other many-body methods employing the same interaction, \emph{in cases where those methods are reliable}.
It is useful to benchmark in the oxygen isotopic chain, where there are published results from several different many-body methods and no complications from deformation.
As illustrated in Fig.~\ref{fig:GroundStates}(c), the shell-model IM-SRG approach, with a core reference state (dashed line), leads to increasing error as valence particles are added.
At harmonic-oscillator shell closures, changing the core leads to a large change in the ground-state energy.
ENO corrects this deficiency, where energies agree well with other methods throughout the chain.
Similar improvement is seen in the calcium and nickel isotopic chains with a few notable discrepancies.
An analogous pattern was seen in deformed neon isotopes \cite{Stroberg2016}, where the multi-reference IM-SRG appeared to decouple a spherical excited state.
For the nickel isotopes, we stop at $^{74}$Ni as the valence-space diagonalization becomes expensive (though not infeasible), but techniques such as importance-truncation \cite{Stumpf2016} provide a clear path forward.

The valence-space approach is not restricted to the vicinity of shell closures;
to demonstrate this versatility, we present the carbon, nitrogen, and sodium isotopic chains in Figs.~\ref{fig:GroundStates}(a), \ref{fig:GroundStates}(b), and \ref{fig:GroundStates}(d), respectively.
The resulting ground-state energies agree well with those obtained with other many-body methods \cite{Jansen2014}, though
the binding energy of $^{12}$C differs significantly from those obtained with IT-NCSM and multi-reference IM-SRG.
While some discrepancy should be due to the normal-ordering approximation, we leave a more thorough investigation for future work. No previous ab initio results exist for sodium isotopes.

\begin{figure}
\includegraphics[width=\columnwidth]{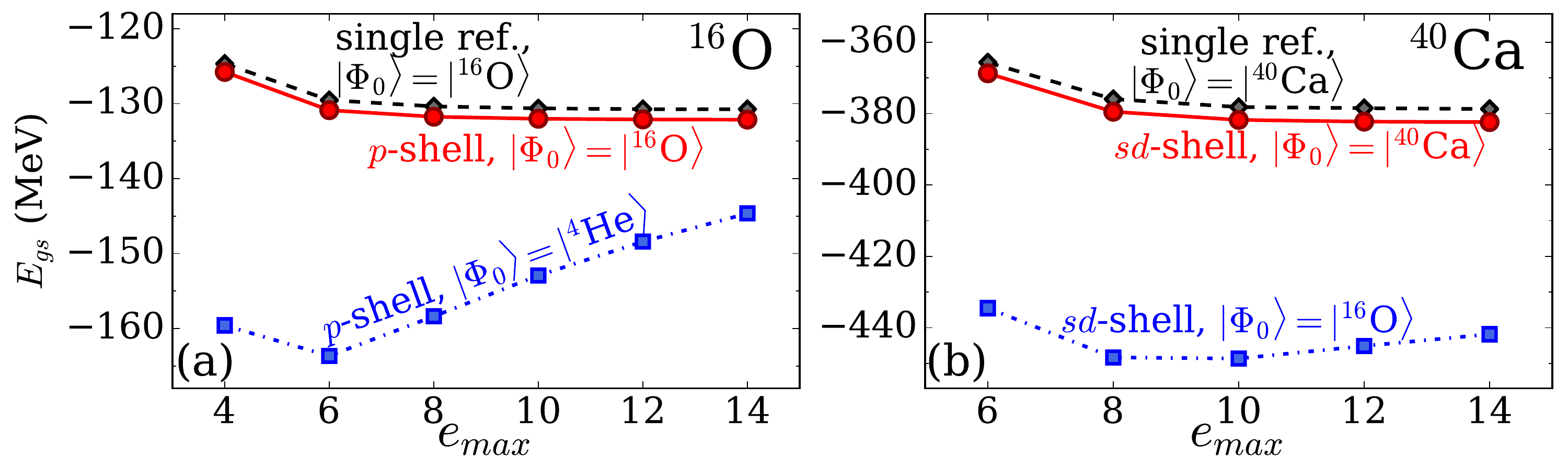}
\caption{(a) Convergence of the $^{16}$O ground-state energy as a function of $e_{\mathrm{max}}$. The curve labeled ``single ref.'' is obtained by decoupling a single Slater determinant representing the ground state. 
The other curves are obtained by decoupling the $0p$ shell, using either a $^{4}$He or $^{16}$O Slater determinant as the normal-ordering reference $|\Phi_0\rangle$. (b) The analogous case for $^{40}$Ca in the $sd$ shell.}
\label{fig:OxygenConvergence}
\end{figure}

An important issue arising in the upper $p$-shell, illustrated in Fig.~\ref{fig:OxygenConvergence}(a) for $^{16}$O, is that shell-model IM-SRG calculations (using a $^{4}$He reference) do not converge with $e_{\mathrm{max}}$, while ENO calculations do.
Similar behavior is observed for other upper-$p$-shell nuclei.
The most likely reason is that the Hartree-Fock single-particle wave functions for the $^{4}$He and $^{16}$O reference states are quite different. Since the former are not optimized for $^{16}$O, we expect that three- and higher-body operators induced by the IM-SRG flow are necessary for a proper description; their omission leads to the observed lack of convergence.
%
%
This effect likely arises in other methods deriving shell-model interactions \cite{Lisetskiy2008,Dikmen2015,Jansen2016} and, as shown in Fig.~\ref{fig:OxygenConvergence}(b), is also present in the upper $sd$-shell, albeit more weakly.

Calculations of $^{16}$O and $^{40}$Ca provide a check on a major source of uncertainty in the valence-space approach since they consist of a single Slater determinant in the $p$ and $sd$ shells, respectively. 
The valence-space results should be identical to those of the single-reference IM-SRG, and any discrepancy must be due to the additional valence-space decoupling.
If this decoupling were perfectly unitary, it would not cause any error.
However, the IM-SRG(2) approximation spoils unitarity, and a small error (approximately 1\%) arises due to this extra decoupling step.

\begin{figure}
\includegraphics[width=\columnwidth]{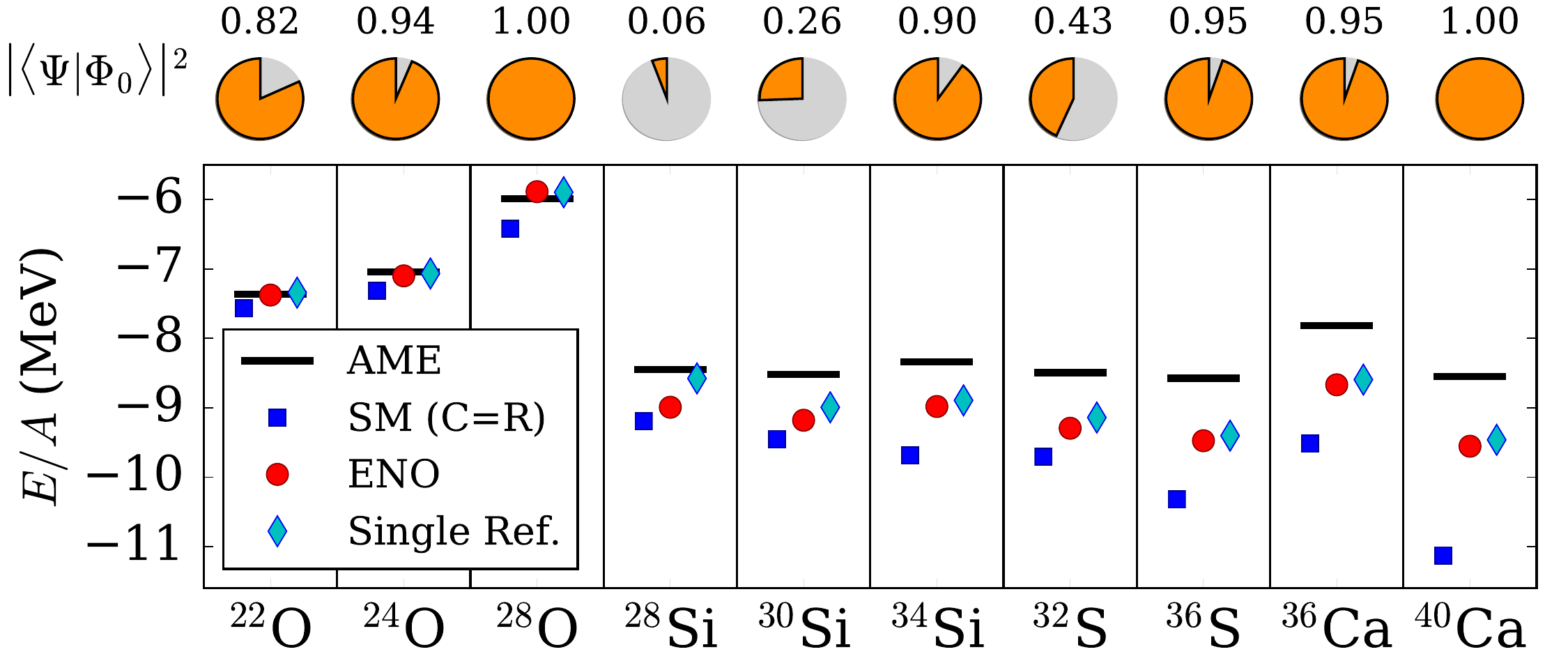}
\caption{Ground-state energy per nucleon of closed $sd$-shell nuclei, calculated with the shell-model IM-SRG using the core as the reference SM(C=R), the ENO approach (here equivalent to TNO), and single-reference IM-SRG. The black bars indicate the experimental values \cite{Wang2012}, and the orange circles indicate the overlap of the ENO ground-state wave function $|\Psi\rangle$ with the reference determinant $|\Phi_0\rangle$.}
\label{fig:ClosedShells}
\end{figure}

In Fig.~\ref{fig:ClosedShells}, we show the ground-state energy per nucleon for several closed $p$- and $sd$-shell nuclei, where the ENO results agree well with the large-space results.
The circles above Fig.~\ref{fig:ClosedShells} indicate the overlap of the reference determinant with the valence-space ground state.
The small overlap for $^{28}$Si indicates that the reference is a poor approximation of the ground state, suggesting that the single-reference calculation selects an excited $0^{+}$ state.

\begin{figure}
\includegraphics[width=1.0\columnwidth]{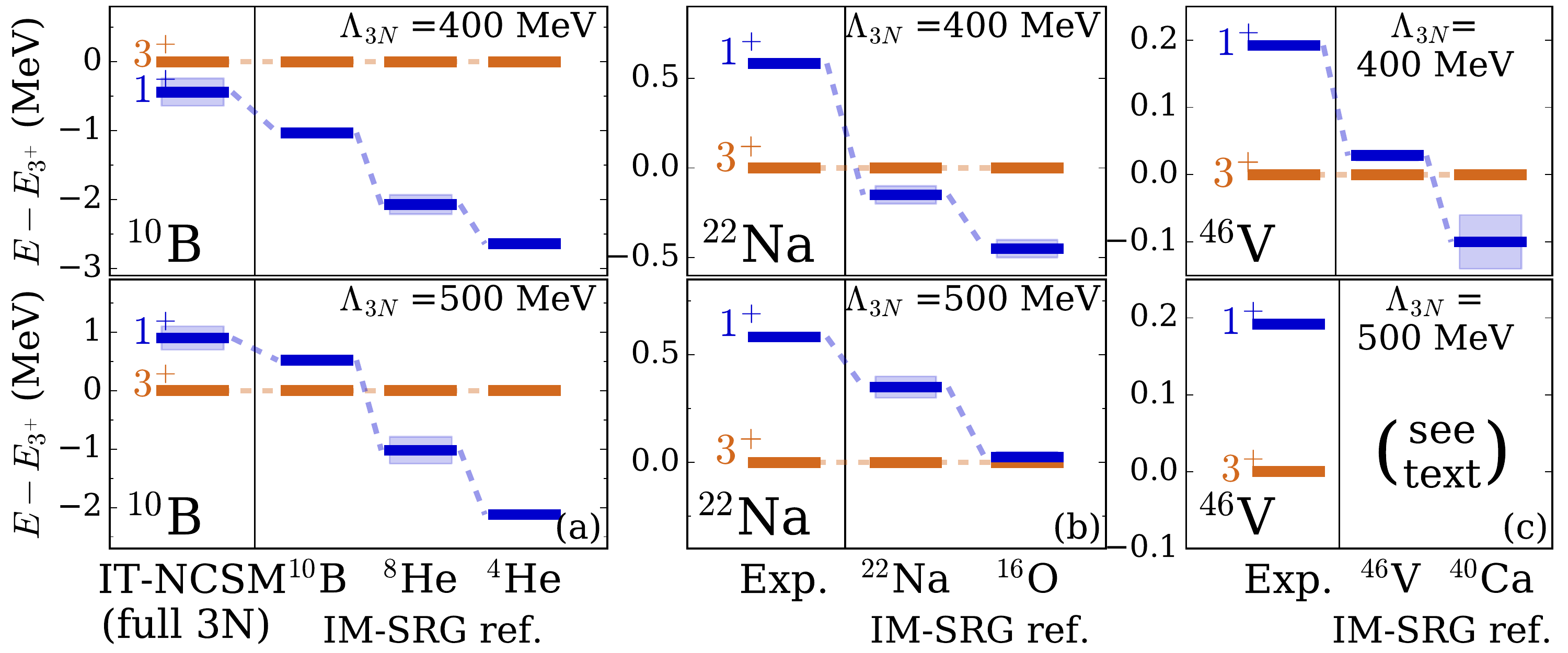}
\caption{(a) Energies of the lowest $1^{+}$,$3^{+}$ states of $^{10}$B, calculated with two different cutoffs for the 3N force, $\Lambda_{\mathrm{3N}}$=400,500 MeV, using both IM-SRG and IT-NCSM. The shaded bands indicate the estimated uncertainty due to model space extrapolations. For the IM-SRG calculations, results for different references $|\Phi_{0}\rangle$ are given. The same is shown for (b) $^{22}$Na and (c)$^{46}$V, comparing IM-SRG to experiment.}
\label{fig:B10}
\end{figure}

Finally, 3N forces are important for spectroscopy. 
A famous case is that of $^{10}$B, where 3N forces are necessary to reproduce the experimental ground-state spin \cite{Navratil2002,*Caurier2002,Pieper2002}.
As shown in Ref.~\cite{Gebrerufael2015}, given an adequate reference, the normal-ordered two-body approximation sufficiently captures this physics.
Fig.~\ref{fig:B10}(a) shows results obtained with various references compared to IT-NCSM results, which include full 3N forces. The two choices of the 3N cutoff demonstrate the sensitivity of this level ordering to the details of the 3N force, and the relevant physics is captured in the ENO calculations.

The analogous $sd$- and $pf$-shell systems are $^{22}$Na and $^{46}$V, respectively.
Results for the $1^+_1$/$3^+_1$ energy splittings in these nuclei are shown in Figs.~\ref{fig:B10}(b) and \ref{fig:B10}(c) for two choices of reference.
As these nuclei are not within reach of IT-NCSM or other large-space methods, we compare to experiment, where we observe a similar effect to that in $^{10}$B.
For $^{46}$V, the sizeable SRG-induced many-body forces for the $\Lambda_{\mathrm{3N}}\!=\!500\MeV$ interaction lead to unreliable results \cite{Roth2012}, so we report only the $\Lambda_{\mathrm{3N}}\!=\!400$~MeV result.
To our knowledge, these are the first ab initio calculations to reproduce the experimental 1$^{+}_1$/3$^{+}_1$ ordering in these systems.

In conclusion, we have generalized the IM-SRG framework to ensemble reference states, allowing approximate inclusion of residual 3N forces in the valence space.
Results agree with other large-space ab initio methods to the 1\% level, but now extend to ground and excited states of essentially all light and medium-mass nuclei, including deformed systems.
In the case of the upper $p$ and $sd$ shells, the ENO approach is required to obtain converged results.
For the specific cases of $^{10}$B, $^{22}$Na, and $^{46}$V, where residual 3N forces are essential to obtain the correct $1^+_{1}/3^{+}_1$ ordering, we have shown that this approach captures the relevant physics.
The unique combination of accuracy, versatility, and low computational cost establishes the valence-space IM-SRG as a powerful ab initio tool for addressing fundamental questions in nuclei.

\begin{acknowledgments}
S.~R.~S would like to thank T.~D.~Morris and N.~Parzuchowski for enlightening discussions, and K.~G.~Leach for providing additional computational resources.
The IM-SRG code used in this work makes use of the Armadillo \texttt{C++} library \cite{Armadillo}.
TRIUMF receives
funding via a contribution through the National Research
Council Canada. This work was supported in part by
NSERC, the NUCLEI SciDAC Collaboration under
U.S.~Department of Energy Grants No.~DE-SC0008533 and
DE-SC0008511, the National Science Foundation under
Grant No.~PHY-1404159, the European Research Council
Grant No.~307986 STRONGINT, the Deutsche Forschungsgesellschaft
under Grant SFB 1245, and the BMBF under Contract No.~05P15RDFN1.
Computations
were performed with an allocation of computing
resources at the J\"ulich Supercomputing Center,
Ohio Supercomputer Center (OSC), and the Michigan
State University High Performance Computing Center
(HPCC)/Institute for Cyber-Enabled Research (iCER).
\end{acknowledgments}

\bibliographystyle{apsrev4-1}
\bibliography{library}{}
\end{document}